\documentclass[12pt,a4paper]{article}
\usepackage{cite}
\usepackage[dvips]{graphicx,color}

\begin{document}
\begin{sloppypar}

\title{Measurement of ${\rm p} {\rm d} \rightarrow {\rm ^3He}
\hspace{0.2em} \eta$ in the $S_{11}$ Resonance}

\author{GEM Collaboration\\ \\M. Betigeri$^{\ a}$, J. Bojowald$^{\
b}$, A. Budzanowski$^{\ c}$, A. Chatterjee$^{\ a}$,\\ M. Drochner$^{\
d}$, J. Ernst$^{\ e}$, S. F\"{o}rtsch$^{\ b,}$\thanks{also at National
Accelerator Centre, Faure, South Africa}, L. Freindl$^{\ c}$, D.
Frekers$^{\ f}$,\\ W. Garske$^{\ f}$, K. Grewer$^{\ f}$, A.
Hamacher$^{\ b}$, S. Igel$^{\ b}$, J. Ilieva$^{\ b,g}$, R. Jahn$^{\
e}$,\\ L. Jarczyk$^{\ h}$,  G. Kemmerling$^{\ d}$, K. Kilian$^{\ b}$,
S. Kliczewski$^{\ c}$, W. Klimala$^{\ b,h}$,\\ D. Kolev$^{\ i}$, T.
Kutsarova$^{\ g}$, J. Lieb$^{\ j}$, G. Lippert$^{\ b}$, H. Machner$^{\
b}$\thanks{e-mail: h.machner@fz-juelich.de}, \\A. Magiera$^{\ h}$, H.
Nann$^{\ b}$\thanks{on leave from IUCF, Bloomington, Indiana, USA}, L.
Pentchev$^{\ g}$, H. S. Plendl$^{\ k}$, D. Proti\`c$^{\ b}$, \\B.
Razen$^{\ b,e}$, P. von Rossen$^{\ b}$, B. R. Roy$^{\ a}$, R.
Siudak$^{\ c}$,J. Smyrski$^{\ h}$,\\ A. Strza{\l}kowski$^{\ h}$, R.
Tsenov$^{\ i}$, P. A. {\.Z}o{\l}nierczuk$^{\ b,h}$, K. Zwoll$^{\ d}$ \\
\vspace {0.5cm}
\\
\\ \small
\noindent {\it a. Nuclear Physics Division, BARC, Bombay, India}
\\ \small
\noindent {\it b. Institut f\"{u}r Kernphysik, Forschungszentrum
J\"{u}lich, J\"{u}lich, Germany}
\\ \small
\noindent {\it c. Institute of Nuclear Physics, Krakow, Poland}
\\ \small
\noindent {\it d. Zentralinstitut f\"{u}r Elektronik,
Forschungszentrum J\"{u}lich, J\"{u}lich, Germany}
\\ \small
\noindent {\it e. Institut f\"ur Strahlen- und Kernphysik der
Universit\"at Bonn, Bonn, Germany}
\\ \small
\noindent {\it f. Institut f\"ur Kernphysik,  Universit\"at
M\"unster, M\"unster, Germany}
\\ \small
\noindent {\it g. Institute of Nuclear Physics and Nuclear Energy,
Sofia, Bulgaria}
\\ \small
\noindent {\it h. Institute of Physics, Jagellonian University,
Krakow, Poland}
\\ \small
\noindent {\it i. Physics Faculty, University of Sofia, Sofia,
Bulgaria}
\\ \small \noindent {\it j. Physics Department, George Mason University,
Fairfax, Virginia, USA}
\\ \small \noindent {\it k. Physics Department, Florida State University,
Tallahassee, Florida, USA}}

\date{ }
\maketitle

\begin{abstract}
We have measured the reaction ${\rm p} {\rm d} \rightarrow {\rm
^3He} \hspace{0.2em} \eta$ at a proton beam energy of 980 MeV,
which is 88.5 MeV above threshold using the new ``germanium wall''
detector system. A missing--mass resolution of the detector system
of 2.6$\%$ was achieved. The angular distribution of the meson is
forward peaked. We found a total cross section of (573 $\pm$ 83
(stat.) $\pm$ 69 (syst.) ) nb. The excitation function for the
present reaction is described by a Breit Wigner form with
parameters from photoproduction.
\end{abstract}

The production of $\eta$--mesons is interesting because it opens
the possibility of studying the interaction between the lightest
isoscalar particle and the nuclear environment. Haider and Liu
were the first to show that even bound $\eta$-nucleus systems,
i.e. $\eta$--mesic nuclei, could be possible \cite{haider1}. Based
on the results of Bhalerao and Liu \cite{bhalerao} they found an
attractive $\eta$--N interaction which in their calculations leads
to bound states for nuclei with mass number A $\ge$ 10
\cite{haider2}. Rakityanski et al. \cite{rakityanski} even relaxed
this condition to A $\geq$ 2. The widths of such states were
predicted to be narrow enough to be observable for nuclei with A
$\geq$ 4. Wycech et al. \cite{wycech} also predicted the formation
of mesic nuclei in ${\rm d} {\rm d} \rightarrow {\rm ^4He}
\hspace{0.2em} \eta$, but not in ${\rm p} {\rm d} \rightarrow {\rm
^3He} \hspace{0.2em} \eta$. In contrast, Abaev and Nefkens
\cite{abaev}, as well as Wilkin \cite{wilkin} showed that the
formation of quasi--bound $\eta$--${\rm ^3He}$ states in the
reaction ${\rm p} {\rm d} \rightarrow {\rm ^3He} \hspace{0.2em}
\eta$ should indeed be possible.

In addition, the reaction ${\rm p} {\rm d} \rightarrow {\rm ^3He}
\hspace{0.2em} \eta$ is of interest due to its surprisingly large
cross section close to threshold making this reaction a prime
candidate for the source of $\eta$--mesons in tagged
$\eta$--facilities \cite{mayer}.

\begin{figure}[htbp]
  \begin{center}
\includegraphics[width=8cm]{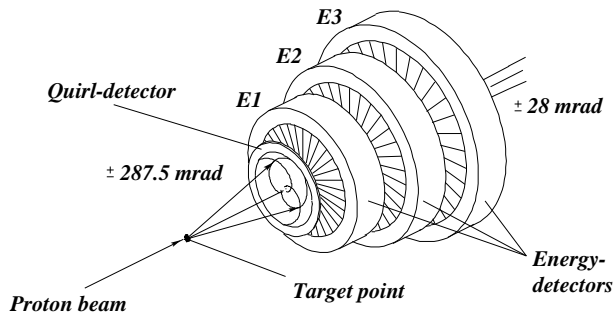}
  \caption{The Germanium Wall \small The detector system ``germanium wall''.
    In the present measurement detector E2 was removed.}
  \end{center} 
\label{GeWall}
\end{figure}

A detector system called the ``germanium wall'' \cite{hardware}
was built at the COSY facility in Julich (see Figure
\ref{GeWall}).
In its complete setup, the germanium wall is a stack of four
position sensitive high--purity germanium detectors having a
conical acceptance with an opening angle of $\pm 287.5$ mrad. In
the centre of each detector is a hole with a size of $\pm 28$ mrad
allowing the primary beam to pass through. Two types of detectors
are used, one 1.3 mm thin diode (``quirl--detector'') for
determining the reaction vertices through its good position
resolution given by the crossing of two counterrotating spirals
and three 17 mm thick diodes for measuring the particle energies
(``energy--detectors''). For further details see Ref.
\cite{hardware}. The setup used for the present measurement
consisted of one quirl and two energy--detectors (Quirl, E1 and
E3, see Figure \ref{GeWall}).
\begin{figure}[htbp]
  \begin{center}
 \includegraphics[width=8cm]{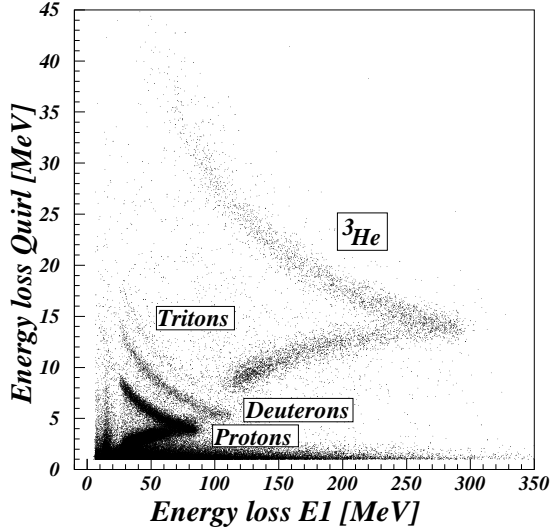}
   \end{center} \vspace{-0.5cm}
  \caption[The Germanium Wall]{\label{DeltaEE} \small Particle identification
    using a $\Delta$E--E spectrum measured with the first two detectors of
    the ``germanium wall''. Protons, deuterons, tritons and ${\rm ^3He}$--particles
    can be identified as indicated in the figure.}
\end{figure}

First measurements with the ``germanium wall'' showed the good
missing--mass resolution of the system. The reaction ${\rm p} {\rm
d} \rightarrow {\rm ^3He} \hspace{0.2em} \eta$ was studied at a
proton beam energy of 980 MeV (88.5 MeV above threshold) leading
to almost 4$\pi$ acceptance of the detector system for the product
${\rm ^3He}$--particles. We performed two runs at different times.

The target was a cell filled with liquid deuterium with 6 mm
diameter and thicknesses of 2.4$\pm$0.2 mm (run A) and 4.4$\pm$0.2
mm (run B), respectively \cite{nake}. The COSY extracted proton
beam was focussed onto the target yielding a spot with a radius
$\sigma=0.5\ mm$  and a divergence of 6 mrad. These parameters
together with the short distance between target to detector yields
a total angular uncertainty of 16 mrad, where the individual
contributions are linearly added. This uncertainty is much larger
than that resulting from the position resolution of the detector,
which is in the order of 2 mrad. The beam had a momentum spread of
$\Delta p/p=8\times 10^{-4}$ \cite{Bet99a}.

\begin{figure}[htbp]
  \begin{center}
    \includegraphics[width=10cm]{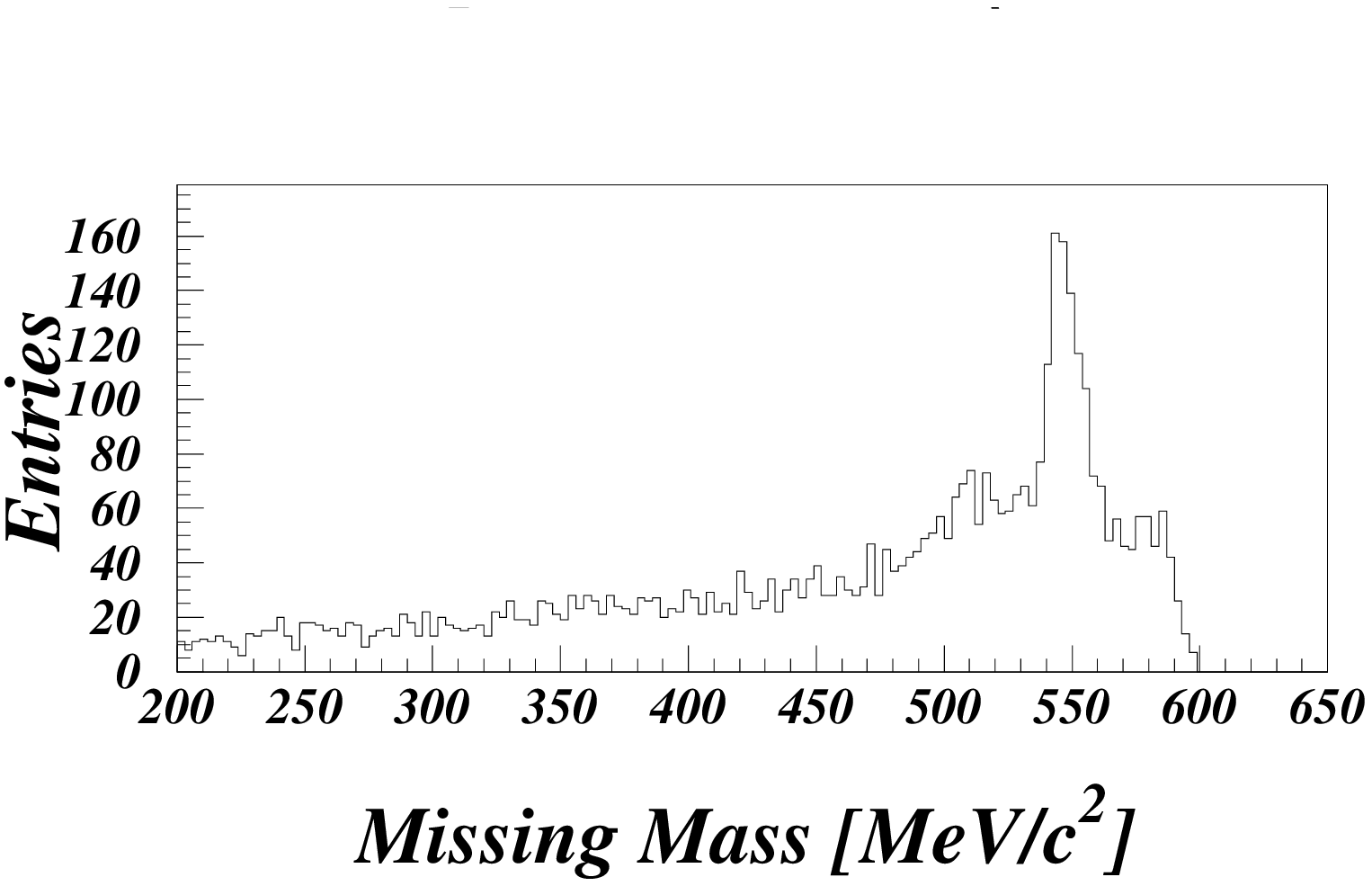}
    \vspace{-0.5cm}
  \end{center} \vspace{-0.5cm}
  \caption[Missing Mass]{\label{MMSep97} \small Missing--mass spectrum. The $\eta$--peak
    at approx. 547 MeV/${\rm c}^2$ is visible with a resolution of (14.2 $\pm$ 1.2)
    MeV/${\rm c}^2$ FWHM. The background is caused by multi--pion production.}
\end{figure}

The energy and direction of the emerging ${\rm ^3He}$--particles
were measured by the ``germanium wall''. Figure \ref{DeltaEE}
shows a $\Delta$E--E spectrum demonstrating the capability of the
detector system for particle identification. Through energy and
emission direction measurement of ${\rm ^3He}$--particles, the
missing--mass was calculated. A missing--mass spectrum for run B
is shown in Figure \ref{MMSep97}. The $\eta$--peak is clearly
visible with a resolution of $\sigma=(6.1 \pm 0.5) MeV/c^2$.
Background is mainly caused by multi--pion production (e.g. ${\rm
p} {\rm d} \rightarrow {\rm ^3He} \hspace{0.2em} \pi^+ \pi^-$,
${\rm p} {\rm d} \rightarrow {\rm ^3He} \hspace{0.2em} \pi^0
\pi^0$, etc.). Low--energy ${^3He}$ background events at small
angles were not detected, because of the minimum opening of the
detector system. These events correspond to large relative
energies between the pions and thus to large missing--mass values.
Therefore, the spectrum is truncated at 600 MeV/${\rm c}^2$. For
run A a missing--mass resolution of $\sigma=(8.2 \pm 0.6) MeV/c^2$
is obtained. The whole body of data was divided into 5 and 6
angular bins for run A and run B, respectively. For each bin a
Gaussian together with a background function were fitted to the
corresponding missing--mass spectrum above 450 $MeV/c^2$. For run
B, two different shapes for the background were assumed: a
polynomial of third order and a function
$BG=\sqrt{a_0\left[1-\left(\frac{mm-a_1}{a_1}\right)^2\right]}
\frac{mm^{a_2}}{a_3}$ with $mm$ the missing--mass and $a_i$
parameters to be fitted. Both functions lead to the same results.
For run A, the background is not so clearly separated for large
missing--masses from the $\eta$ peak as it is in run B. Therefore,
several functions were tested. Polynomials were fitted to the
range lower than the $\eta$ peak and only the first point above
the peak. Alternatively a step like function (cummulative
Lorentzian) was fitted to all data. The $\eta$ peak was always
assumed to be a Gaussian.  Finally the number of $pd\rightarrow
{^3He}\eta$ events was obtained by integrating the Gaussians and
weighted means were deduced. Further details of the data analysis
procedure are given elsewhere \cite{Gre99}.

\begin{figure}[htbp]
  \begin{center}
\includegraphics[width=10cm]{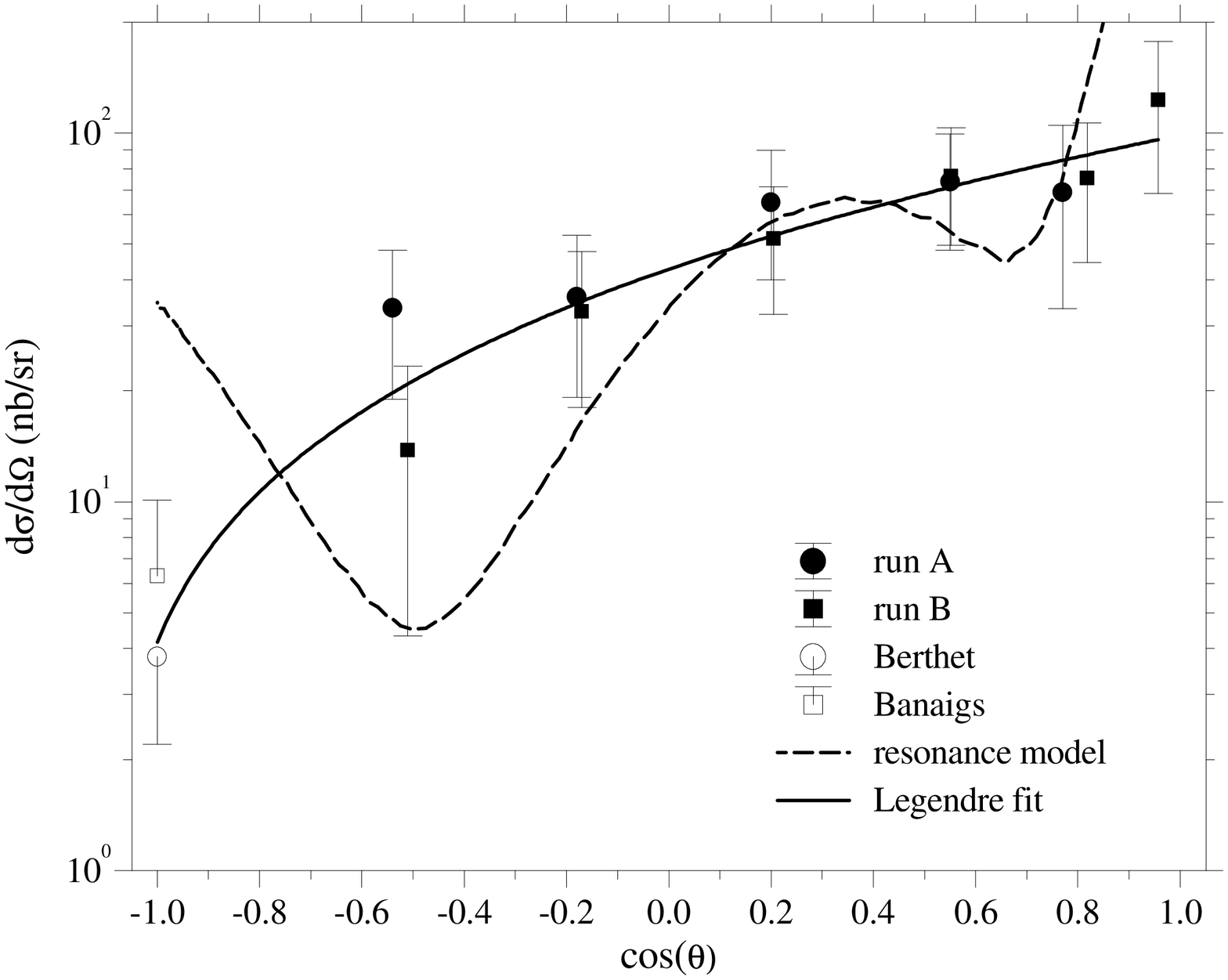}
  \end{center} \vspace{-0.5cm}
  \caption[Angular Distribution]{\label{AngularDistribution}
  \small Angular distribution for the reaction ${\rm p} {\rm d}
  \rightarrow {\rm ^3He} \hspace{0.2em} \eta$ at 88.5 MeV above
  threshold in the centre of mass system. The solid curve shows a
  Legendre polynomial fit to the data, the dashed curve a
  calculation within a resonance model (see text).}
\end{figure}

Due to beam halo during the experiments the intensity of the beam
had to be reduced to a level of $10^5$ protons per second. Thus
pile up and detector damage were avoided, but event statistics was
strongly reduced. The measured angular distribution is shown in
Figure \ref{AngularDistribution}.
The error bars shown represent the statistical errors only. In
addition, there are systematic uncertainties: target thickness
$10\%$ and $5\%$ in the different runs, respectively, luminosity
calibration $7\%$, corrections due to trigger and detector
inefficiencies due to nuclear interactions in Germanium (see Ref.
\cite{Mac99}) $5\%$. The total systematic error of $13\%$ and
$10\%$ (when added in quadrature) for the two runs, respectively
is smaller than the statistical error. The efficiency of the data
analysis ($80\%$) was studied by Monte Carlo simulations. The
simulated detector response was found to be in excellent agreement
with the experiment \cite{Gar99}. Before and after each run the
detector was checked with radioactive sources. No significant
deviations in the amplifier gains were found as expected, since
the electronic circuits were kept at a constant temperature. Since
both runs were performed under different experimental conditions
with respect to beam halo, target thickness and distance between
target and detector \cite{Gre99}, different systematic errors lead
to an enhancement of run A compared to run B which is slightly
above the statistical error given above. The halo was 2.2 times
more intense during run A than in run B thus leading to a larger
combinatorial background than in run B and hence to larger error
bars of the integrated Gaussians for the $\eta$ peak.

The angular distribution is forward peaked. This is in contrast to
the results of Mayer et al. \cite{mayer} reporting almost
isotropic distributions close to threshold. The cross sections of
Banaigs et al. \cite{banaigs} measured through ${\rm d} {\rm p}
\rightarrow {\rm ^3He} \hspace{0.2em} \eta$ at a slightly higher
excitation energy (corresponding to an equivalent proton kinetic
energy of 1047 MeV) agree with our data. When we fit Legendre
polynomials to the present data, an unphysical negative value for
$\cos (\theta )=-1$ is obtained. In order to overcome this
deficiency we have added one data point at $\cos(\theta)$=-1 from
Ref. \cite{berthet}, who had measured an excitation function for
this angle. This point is also shown in Fig.
\ref{AngularDistribution}. In order to take into account the fact
that the point was measured at a slightly different energy of $0.5
\%$, we have doubled its statistical error. Another data point in
the literature taken at a somewhat higher beam energy (see Ref.
\cite{banaigs}) is also shown in Fig. \ref{AngularDistribution}.
Again its error bar was doubled. A Legendre polynomial fit to all
points of the angular distribution, i.e. the present ones from
both runs as well as the one from  Ref.'s \cite{berthet,banaigs}
weighted by their total errors, yielded parameters $A_0=45.6 \pm
5.9,\ A_1= 47.3\pm 10.9$, and $A_2=5.8\pm 8.5$ all given in nb/sr,
and with a $\chi ^2/nfree =0.3$ from which a total cross section
of (573 $\pm$ 74) nb follows. In addition to this statistical
error the systematic error is assumed to be 69 nb. The result is
insensitive to the added point, because the differential cross
section is small for the backward angle emission. Including higher
degrees in the fit procedure does not improve the fit, a lower
degree gives an unphysical negative value for $\cos (\theta )=-1$.

Kingler \cite{kingler} has extended a model originally developed
for pion production \cite{Harz}  to include higher--resonances.
The original model was limited to only pion exchange, $\Delta$
resonance excitation as well as non resonant contributions. The
extension treats other meson exchanges as well as higher nucleon
resonances than the $\Delta$. The vertex function for the
different baryon--baryon--meson couplings were calculated in a
simple quark model and are momentum dependent. For the present
reaction the largest contribution to the cross section comes from
the $NN\rho$ and $NN\omega$ interactions while the contribution
due to $NN^*(1535)\pi$ interaction is one order of magnitude
smaller. The contributions of other resonances like $N^*(1440)$,
$N^*(1650)$, and $N^*(1710)$ are even smaller. However, the form
factor is calculated only for harmonic oscillator wave functions
with the frequency being a free parameter varied to fit the
experimental data. The model predictions are shown in Fig.
\ref{AngularDistribution} as dashed curve. Obviously, the
calculation shows structures not observed in the present data.

\begin{figure}[htbp]
  \begin{center}
 \includegraphics[width=10cm]{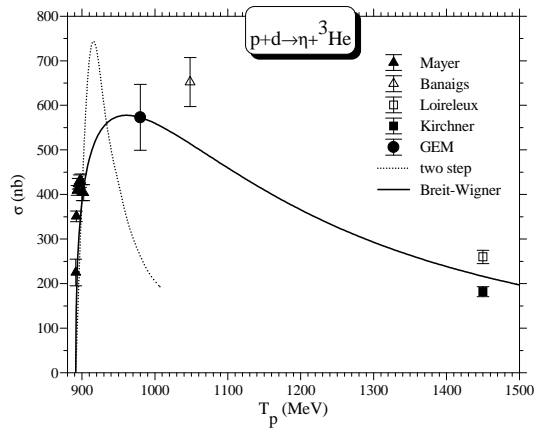}
  \end{center} \vspace{-0.5cm} \caption[Angular
Distribution]{\label{exfunc} \small Total cross sections for
$pd\rightarrow {\rm ^3He}\eta$ reaction as function of the proton beam
energy. Data are indicated by different symbols. The  kinematical
velocity matching in a two step process is shown as dotted curve. The
solid curve is a calculation employing the matrix element from
photoproduction on the proton.}
\end{figure}

The extracted total cross section together with earlier data from
Mayer et al. \cite{mayer} is shown in Figure \ref{exfunc}. From
the angular distributions given by Banaigs et al. \cite{banaigs},
Loireleux \cite{Loireleux} and Kirchner \cite{Kirchner} we
extracted total cross sections by fitting Legendre polynomials.
The results are also shown in Figure \ref{exfunc}. Kirchner
claimed that the data from Ref. \cite{Loireleux} suffer from
important electronic problems. No further details are given. The
data from Mayer et al. \cite{mayer} indicate a strongly rising
cross section close to threshold.

Also shown in the Figure is a normalized calculation within a
two--step model developed by Kilian and Nann \cite{Kil90}. Within
this model a pion is produced in a first step through
$pp\rightarrow d\pi^+$. In a second step, this pion produces the
$\eta$ in an interaction with the neutron. A kinematical velocity
matching yields the maximum close to threshold. The present data
do not support this model as the dominant reaction mechanism.

 The energy region from 900--1100 MeV corresponds to the centre of the $N^\ast$ $S_{11}$
resonance ($\Gamma$ $\sim$ 200 MeV) known to couple strongly to
the $\eta$--$N$ channel \cite{PDG99}. One may therefore attempt to
describe the cross section by an intermediate $N^\ast$(1535)
resonance excitation. The cross section is calculated as
\begin{equation}\label{cross_sect}
  \sigma(E)=\frac{p_\eta}{p_p}|M(E)|^2
\end{equation}
with $E$ the excitation energy and $M$ the matrix element. All
momenta $p$ are in the centre of mass system. This is calculated
as in photoproduction on the proton \cite{krusche}
\begin{equation}\label{Breit_wigner}
  |M(E)|^2=\frac{A\Gamma_R^2}{(E-m_r)^2+\Gamma(E)^2}
\end{equation}
with
\begin{equation}\label{gamma}
  \Gamma(E)=\Gamma_R\left(b_{\eta} \frac{p_{\eta}}{p_{\eta,R}}+b_{\pi}
  \frac{p_{\pi}}{p_{\pi,R}}+b_{\pi\pi}\right).
\end{equation}
Similar to Ref. \cite{krusche}, we applied a width at the
resonance of $\Gamma_R$=200 MeV, a Breit-Wigner mass of $m_R$=1540
MeV/c$^2$. The branching ratios were set to $b_{\eta}$=0.47 for
the $\eta$ decay, $b_{\pi}$=0.48 for the pion decay and
$b_{\pi\pi}$=0.05 for the two pion decay \cite{PDG99}. The momenta
at the resonance position are indicated by the index $R$. The only
free parameter is the strength $A$ taken to be 241 nb in order to
fit to the present data point. The calculation is shown in Figure
\ref{exfunc} as a solid curve. The trend of the data is
reproduced, which may be taken as an indication that production of
the $N^\ast$(1535) resonance is the dominant reaction mechanism
and that the product of kinematics and form factor changes only
very little over the present energy range. The overall shape of
the calculation slightly underestimates the data of Mayer et al.
\cite{mayer} close to threshold. For a more detailed investigation
of fine structure, additional data in this region is needed.
 An enhancement close to threshold was also seen in $\eta$
production in NN interactions \cite{Cal96,Cal98} and was attributed to
a strong final state interaction. The agreement between the excitation
functions for $pd\rightarrow {\rm ^3He}\eta$ and $\gamma p\rightarrow
p\eta$ reactions excludes strong FSI between the nucleus and the $\eta$
except for the near threshold region.

We gratefully acknowledge the COSY crew for their efforts
providing us with a good beam. We are thankful for support by BMBF
Germany (06 MS 882), Internationales B\"{u}ro des BMBF, SCSR Poland
(2P302 025 and 2P03B 88  08),  NATO Scientific Affairs, and COSY
J\"{u}lich.


\end{sloppypar}
\end{document}